\documentstyle[aas2pp4,lscape]{article}
\lefthead{Alcock et al.}
\righthead{Difference Image Analysis}

\begin{document}
\title{Difference Image Analysis of Galactic Microlensing}

\begin{center}
{\large\bf II. Microlensing Events}
\end{center}

\author{C. Alcock\altaffilmark{1,2}, R.A. Allsman\altaffilmark{3}, D. Alves\altaffilmark{1,4},
T.S. Axelrod\altaffilmark{5}, A.C. Becker\altaffilmark{6}, D.P. Bennett\altaffilmark{1,2},\\
K.H. Cook\altaffilmark{1,2}, A. J. Drake\altaffilmark{5}, K.C. Freeman\altaffilmark{5}, K. Griest\altaffilmark{2,7},
M.J. Lehner\altaffilmark{8}, S.L. Marshall\altaffilmark{1,2},\\ D. Minniti\altaffilmark{1,13}, 
B.A. Peterson\altaffilmark{5}, M.R. Pratt\altaffilmark{9}, P.J. Quinn\altaffilmark{10},
C.W. Stubbs\altaffilmark{2,5,6},\\ W. Sutherland\altaffilmark{11},
A. Tomaney\altaffilmark{6}, T. Vandedei\altaffilmark{7}, and D.L. Welch\altaffilmark{12}}

\author{\bf(The MACHO Collaboration)}

\altaffiltext{1}{Lawrence Livermore National Laboratory, Livermore, CA 94550} 
\altaffiltext{2}{Center for Particle Astrophysics, University of California, Berkley, CA 94720} 
\altaffiltext{3}{Supercomputing Facility, Australian National University, Canberra, ACT 0200, Australia}
\altaffiltext{4}{Department of Physics, University of California, Berkeley, CA 95616}
\altaffiltext{5}{Mount Stromlo and Siding Spring Observatories, Weston Creek, Canberra, ACT 2611, Australia}
\altaffiltext{6}{Department of Astronomy and Physics, University of Washington, Seattle, WA 98195}
\altaffiltext{7}{Department of Physics, University of California, San Diego, CA 92093}
\altaffiltext{8}{Department of Physics, University of Sheffield, Sheffield s3 7RH, UK}
\altaffiltext{9}{Center for Space Research, MIT, Cambridge, MA 02139}
\altaffiltext{10}{European Southern Observatory, Karl Schwarzchild Str.\ 2, D-85748 G\"{a}rching bel M\"{u}nchen, Germany}
\altaffiltext{11}{Department of Physics, University of Oxford, Oxford OX1 3RH, UK}
\altaffiltext{12}{Department of Physics and Astronomy, McMaster University, Hamilton, ON L8S 4M1, Canada}
\altaffiltext{13}{Departmento de Astronomia, P. Universidad Cat\'olica, Casilla 104, Santiago 22, Chile}

\begin{abstract}
  The MACHO collaboration has been carrying out Difference Image Analysis
  (DIA) since 1996 with the aim of increasing the sensitivity to the detection
  of gravitational microlensing. This is a preliminary report on the
  application of DIA to galactic bulge images in one field.
  
  We show how the DIA technique significantly increases the number of
  detected lensing events, by removing the positional dependence of
  traditional photometry schemes and lowering the microlensing event
  detection threshold. This technique, unlike PSF photometry, gives the
  unblended colours and positions of the microlensing source stars. We
  present a set of criteria for selecting microlensing events from objects
  discovered with this technique. The 16 pixel and classical
  microlensing events discovered with the DIA technique are presented.
\end{abstract}

\keywords{Cosmology: gravitational lensing - methods: data analysis - 
Galaxy: stellar content; center - stars: brown dwarfs}

\section{INTRODUCTION}

The on-going search for microlensing events in the direction of the galactic
bulge is now reaching its prime with over 300 events detected by the MACHO
(\cite{alc98}), EROS (\cite{bea95}), DUO (\cite{ala97c}) and OGLE
(\cite{pac94}) groups. 

Microlensing research in the bulge, although initially only a test-bed for
the phenomenon of microlensing, has now graduated to become an area for study
in its own right. These microlensing events allow us to study the
atmospheres of clump giants in the bulge (\cite{alc97x}; \cite{hey99}),
and possibly detect the existence of distant extra-solar planets
(\cite{gri98}; \cite{bec98}). The optical depth determined from the
events permit us to calculate the number density and masses of objects
intervening between us and the source stars. The spatial distribution of
these events will help in determining the distribution of mass within our
galaxy. Aside from the benefits from the microlensing to the astronomical
community, there are other important uses of the years of observations of
millions of stars. For instance, within the datasets there are years of
coverage of thousands of variable stars, see for example 
(Alcock et al. 1997c, 1998b).

In this initial Difference Image Analysis (hereafter DIA) study, we once
again use the bulge as a test-bed for microlensing analysis. As before, we
use the bulge as a target because of the large number of microlensing events
observed toward the bulge relative to the LMC. We perform the DIA technique
in an attempt to increase our efficiency for detecting microlensing events.
With the successful result of increasing the detection rate of microlensing
towards the bulge, we can now be optimistic about applying it on data taken
towards the LMC, where the present number of detected events is small.

In the \S 3 we will present the microlensing events found in this analysis.
In \S 4 and \S 5 we will analyse these new results. In the final section we
shall make our concluding remarks about the results and potential of the
technique.

\section{OBSERVATIONS}

This analysis presents results from the application of the DIA technique on 
324 observations of a single $42^{\prime}$ by $42^{\prime}$ field centred 
at $\alpha= 18\arcdeg 01'20''$, $\delta = -28\arcdeg17'39''$ $(J2000)$. 
Observations were taken on the Mount Stromlo and Siding Spring 
Observatories' 1.3m Great Melbourne Telescope with the dual-colour 
4k by 4k {\it Macho camera}. All bulge observations have $150$ second 
exposures. See \cite{alc99a} for further details.

{\it The data analysis strategy and details of the DIA technique are
presented in paper I} (\cite{alc99a}).

\section{RESULTS}

The result of the analysis carried out on three years of bulge data
for an individual field is the detection of thousands of variables,
over a hundred asteroids and 16 probable microlensing events.
Of which nine events are classified as classical events and seven as pixel
lensing events. Classical microlensing involves resolved source
stars. Pixel lensing involves source stars which are too faint or too blended 
to be resolved. A typical difference image is shown on the right of figure 
\ref{typi}, where the reference image is presented on the left.

\placefigure{typi}

Of the $324$ observations put through the bulge DIA pipeline $306$
observations were successfully reduced. This constitutes $\sim80\%$ of the
original data set of observations for this field. The 18 reduction failures
in the analysis occured due to poor data conditions. This included bad 
geometric registration due to saturated registration stars and very 
low transmission relative to the reference image 
(as low as $15\%$ reference image flux).

\subsection{Event Selection}

The search for microlensing events is simplified by the fact that the
template is created by the combination of multiple images over a long time
baseline. Most variables in the constituent images are thus observed at a
range of phases making the reference image close to an average phase for
each variable. This means that a large percentage of variable stars will
have positive and negative excursions from their mean amplitude. However,
microlensing events and asteroids will generally only have excursions above
the baseline. This property allows us to considerably reduce the size of the
search by only investigating results with positive excursions from the
median baseline. This selection means it is quite possible that we could
miss events which have long time scales ($\hat{t} \gtrsim 1$ year).

To perform event selection one must realise that spurious detections of
objects can occur for a number of reasons. The most common sources of these
are asteroids, satellite tracks and cosmic rays. Numerous detections of
cosmic rays occur, but these are removed in the process of matching object
positions in two colours.  Asteroids and satellite tracks are detected in
two colours and are thus more difficult to remove from an individual
observation. To rule out these detections the search for microlensing
detections must be made with at least two consecutive significant
detections in two colours, at any given position. However, one can not
determine the time scale $\hat{t}$ or the amplification {\em A} for
microlensing events which are detected from only two consecutive points
above the baseline in a light curve. Such results can also sometimes be
caused by poor photometry due to bad seeing or CCD defects. For this reason
we searched the photometric database for events of at least 4.5 sigma
significance in at least three consecutive observations. This significantly
lowered the number of short time scale events we expected to detect. For this
data set the cuts led to a microlensing time scale ($\hat t$) lower limit of
approximately five days.

Events with a lower signal-to-noise ratio (hereafter S/N) are selected based
on 4 or more consecutive points. In all cases an event must have a total
S/N of greater than ten in each colour. Each of these so called {\it
point filters} require a positive deviation from baseline (median) 
in consecutive observations of an object. These point filters have
the effect of reducing the number of possible events to 
roughly $15\%$ of the initial data set.

DIA sometimes deals with variable objects of an indeterminate magnitude, so
a set of cuts different from those used previously in Alcock et al. (1997a)
was imposed. Variable stars detected in this analysis were cut in the selection
process with a number of variable star point filters. For example, if there
are four or more consecutive points 3.5 sigma below the median, the result is
deemed to be due to a variable star. Such a result is not consistent with
microlensing and is too significant to be due to poor photometry alone.

Observations of bulge fields are generally only taken once per night
for approximately eight months of the year. This means a large number of
bulge events are missed because they occur during the observing gap. 
A number of possible events are detected as rising at the end of 
the observing season, or falling at the beginning of the next.
But these events can be ambiguous. For such events it is sometimes 
difficult to obtain useful parameters from light curve fitting if 
we do not know the maximum flux of the event. We therefore require 
the maximum amplification occurs during the time the object was monitored. 

In our cuts we do not require the microlensing light curve has a single peak
as is common with many sets of cuts. Not applying this type of cut makes the
removal of variables more difficult, but allows us to detect binary lensing
events where multiple peaks are apparent. Further cuts based on the
microlensing fit were also applied. But these cuts also do not remove binary
events from the list of detected objects (see below). All light curves
passing the first cuts stage were examined before this cut was applied.

With DIA we do not necessarily know the true baseline magnitude of the
objects. So in contrast to Alcock et al. (1997a) we do not impose an
amplitude threshold cut for microlensing events. Furthermore, since DIA is
not affected by source crowding, we made no cut for this either. This
initial set of variable cuts reduced the number of results from $\sim9000$
to 80.

Most of the points in the light curve of a microlensed star should 
lie near the median baseline and be consistent with a microlensing fit.
So the remaining light curves next underwent fitting for microlensing
and for a constant baseline outside the microlensed region. 
A second set of cuts is then performed based on the $\chi^{2}$ values 
from these fits to remove the remaining low amplitude variables.

\placefigure{cut}

We selected events with mircolensing fit $\chi^{2}_{m}, < 2.5$ and constant
baseline fit $\chi^{2}_{c}, < 6$ in the region $t_{0} \pm 2 \hat{t}$.
$\chi^{2}_{m}$ and $\chi^{2}_{c}$ are the reduced chi-squared statistics of
the fits. Microlensing events can vary significantly from a point source
microlensing fit (e.g. events with parallax). For microlensing events we
require $1000/pf \times (\chi^{2}_{c} + \chi^{2}_{m}) < 3.6$, where $pf$ is
flux at peak amplification in ADU. This selection acts as a S/N cut for the
detected events. For events with a peak flux greater than 25000 ADU, we do
not require the microlensing cut or the baseline constancy cut. These events
generally have a high S/N, but may not pass our simple microlensing
$\chi^{2}$ fit cut because of structure. The event 97-BLG-28 (next
section) falls into this category because of the caustic crossing. For these
events we required a more stringent signal to noise cut of $1000/pf \times
(\chi^{2}_{c} + \chi^{2}_{m}) < 1.5$.  The final cut we imposed on all events
was that the microlensing fit colour be greater than $V-R = 0.55$. This
cut removes dwarf novae which can be quite well fitted by a microlensing in
some cases, but are much bluer during outburst.

The effect of these $\chi^{2}$ cuts on the 80 events passing the variable cuts 
is shown in figure \ref{cut}. One can see that most of the events
which fail the constant background cut also fail the microlensing fit cut.
This demonstrates the robustness of this set of cuts.

\subsection{Parameters of Microlensing Events}

The parameters for the sources of candidate microlensing events are 
given in table \ref{pars}, with difference image photometry
colours $(V-R)_{d}$ and those of the nearest matching MACHO database object
$(V-R)_{p}$. The extinction corrected source colours $(V-R)_{c0}$ are also
given. The difference image source coordinates and the separation {\em S} 
of the nearest database object are presented. Events which were alerted on 
are noted.

\placetable{pars}

We have used the Starlink software package {\em Astrom} to perform
astrometry on the coordinates of difference image objects to determine the
RA and Dec of sources. The nearest matching Macho database stars were found
using their coordinates and the results are presented. The associated
errors in {\em S} are mainly due to the uncertainty in the transformation
between difference image template and the pipeline photometry template.

The V-R colours in this table were obtained by the standard Macho 
calibrations (\cite{alv98}). The Cousins magnitudes for the bulge 
are given by:

\begin{equation}
%\small
V_{c} = 23.67 + 0.847 \times B_{m} + 0.156 \times R_{m}
\label{Visual}
\end{equation}

\begin{equation}
%\small 
R_{c} = 23.48 + 0.822 \times R_{m} + 0.182 \times B_{m}
\label{Red}
\end{equation}

Here $B_{m}$ and $R_{m}$ are the Macho camera blue and red passbands,
respectively. Extinctions have been determined from RR Lyrae stars in the
proximity of microlensing events. For RR Lyrae ab stars we have used
$E_{V-R} = (V-R)_{obs} - 0.21$ (\cite{alc98}).  Using the standard
extinction law of Rieke and Lebofsky (\cite{Rie85}), we obtain $A_{v} =
3.96E(V-R)$.

\placefigure{Macho}

As the differential reddening in this field is large
(\cite{alc98}), we chose to extinction correct each event individually
based on the RR Lyrae within $1'$. This technique has some uncertainty 
because of the small numbers of RR Lyrae in each field.
In future we hope to use the more numerous red clump giant stars 
to determine extinctions more accurately. 
The consistency of these two methods has recently been demonstrated 
for Baades window by Alcock et al. (1998b) and Gould et al. 
(\cite{gou98a}).
The difference image colours of the events come from stacking the 
images over the period where the event was two sigma above the 
detection threshold. This gives accurate colours for events which
are independent of any blending in the template. PSF photometry
colours have been determined from the baseline colours of the 
Macho objects. The difference between these colours and the 
DIA colour can be seen in figure \ref{Macho}.
This shows the extent of blending for these sources.
Points are shown with one sigma uncertainties.

\section{EVENTS}

The number of alert events which pass our selection cuts
for the period of this reduction is eight and the number of
new possible microlensing events is also eight.

\placetable{fit}
\placetable{fit2} 

All the events detected in this analysis
were fit with a point source point lens microlensing fit. The values are
given in table \ref{fit} for classical results and in table \ref{fit2} for
pixel lensing.  Plots of the fits are given in figures \ref{Class2} and
\ref{Diff}.  Two binary microlensing events which have not been fit are
given in figure \ref{Unfit}. For most of these events there is
insufficient sampling to accurately determine the source flux. Therefore we
have chosen to use the measured reference image baseline flux as the maximum
flux which can be contributed by the source.  This gave us a limit on the
minimum amplification $A_{min}$ and time scale of the event $\hat{t}_{min}$.
In a future paper (Alcock et al. 1999c) we shall use HST data of the bulge to
determine luminosity function for this and other fields. This will enable us
to determine detection efficiencies for events due to source stars several
magnitudes below the crowding limit of this data.

We shall now briefly discuss the individual microlensing candidates. Events
95-BLG-33, 95-BLG-32 and 95-BLG-38 are classified as excellent microlensing
candidates. Event 95-BLG-21 has an excess of flux at peak amplitude which
could be due to the source passing near a caustic. However, this candidate
is a excelent microlensing candidate event. Event 95-BLG-14 is a
good microlensing candidate, although there are a few missing data
points in the blue passband due to a CCD defect near the source position.
Event 96-BLG-16 has a poor $\chi^{2}$ in the PSF photometry due to the fact
that this source is highly blended. The fit parameters for this event (table
\ref{fit}) are different from those given in figure 7 of paper I as the
baseline flux is fixed for fits in figure \ref{Class}.  In figure 7 of paper I
the source flux has been fit for. Candidate 97-BLG-28 is a poor fit to
single point source microlensing because there is a cusp crossing near
maximum amplification. Still this event is regarded as an excellent microlensing
candidate. A thorough analysis of this event will be presented in a future
paper (\cite{alc99b}).  Event 97-BLG-51 is an example of a moderate signal
lensing event. Although the light curves do not extend far past maximum
light, further points available in the PSF photometry reduction give further
weight. The discrepancy in the two fit amplitudes is due to the fact that
the PSF photometered star is not the lensed star, but rather, is a nearby star
sequestering some of the light in the PSF photometry.

Events 95-BLG-d2, 95-BLG-d3, 95-BLG-d4, 95-BLG-d5, 97-BLG-d1 and 97-BLG-d3
are all good candidate microlensing events. 
Here the use of a {\em d} before the final number means the event was first 
discovered in the difference image analysis. The amplification values
for 95-BLG-d5, 95-BLG-d3 and 97-BLG-d1, are all quite low but it
should be mentioned that these are the minimum possible amplitudes.
Event 96-BLG-d1 is reasonably fit with a microlensing light curve
though it appears to have a feature on the rising side.
This feature could be the signature of a binary microlensing event and 
is considered a good candidate microlensing event. 
Event 97-BLG-d2 is a very poor point source lensing event
but is a very good candidate binary lensing event.
This event will be subjected to a detailed analysis in a future 
paper (Alcock et al. 1999c).

Possible cases of the detection of pixel lensing have been presented in the
past (\cite{cro97}). However, it is believed their survey suffered from
insufficient baseline to rule variable stars out as the sources (Gould \&
Depoy \cite{gou98b}). Of the new DIA results presented here, six are
probably pixel lensing events, based on their nearest resolved star
separations and colours and light curves.  These are 95-BLG-d2, 96-BLG-d1,
97-BLG-d1, 97-BLG-d2, 97-BLG-d3 and 95-BLG-d4. The other two good candidate
events (95-BLG-d3, 95-BLG-d5) are probably classical microlensing events
which fall below the PSF photometry alert system detection threshold
(\cite{alc99b}).

\placefigure{Class}

\placefigure{Class2}

\notetoeditor{These first two figures are part of a single figure and
should span a page each.}

\placefigure{Diff}
\notetoeditor{This figure should also span a page.}

\placefigure{Unfit}

\section{SUMMARY}

We have shown how the DIA technique can be used on archival data to detect
microlensing events undetected with the traditional approach.  This is
achieved by removing the positional constraints and accurately removing
non-variable objects. Nine classical microlensing events and seven
pixel lensing events were found. Of these results, eight events were new events
undetected by the alert system.

DIA is unaffected by unlensed blended flux. Because of this, it is
possible to determine the existence of a blending even where the separation
between source and blended flux centroid is small, by comparing DIA colours
with PSF photometry colours.  Our results prove that it is possible to
determine the source flux from DIA light curves for bright sources and hence
obtain accurate $\hat t$ values for such events. The first pixel lensing
candidates with sufficiently long baselines to prove their existence have
been presented.

Three methods for determining useful parameters from difference image
results have been presented. Firstly for high S/N events, we can fit for the
source flux of the event to obtain $\hat t$. For lower S/N results we can
impose the flux in the reference as a upper bound and obtain a limit on
$\hat t$. Lastly from paper I, we can use a deep colour magnitude diagram
(such as HST) to determine the distribution of source stars colours with
magnitude. With this and the colour of the source, and its associated
uncertainties, we can determine a distribution of allowable $\hat t$
values. This in turn could provide us with a $\hat t$ distribution for the
set of events.

In future, DIA will allow us to gain greater sensitivity to detecting long
time scale low amplification events. The detection of these events will be
accomplished by simply stacking difference images to increase the detectable
signal of these events. For such events parallax can be measured and used
to break the degeneracy in lens mass and position.

We are grateful for the skilled support by the technical staff at Mount
Stromlo Observatory. Work at Lawrence Livermore National Laboratory is
supported by DOE contract W7405-ENG-48. Work at the center for Particle
Astrophysics at the University of California, Berkeley is supported by NSF
grants AST 88-09616 and AST 91-20005. Work at Mount Stromlo and Siding
Spring Observatories is supported by the Australian Department of
Industry, Technology and Regional Development. Work at Ohio State
University is supported in part by grant AST 94-20746 from the NSF. 
W. J. S. is supported by a PPARC Advanced Fellowship. K. G. is grateful 
for support from DOE, Sloan, and Cottrell awards. C. W. S. is grateful 
for support from the Sloan, Packard and Seaver Foundations.
This work was carried out by A.J.D. in partial fulfilment of the 
requirements for the degree of PhD at ANU.

\newpage
\pagestyle{empty}

\begin{figure}
\epsscale{1.00}
%\plotone{f1.ps}
%\figurenum{10}
\caption{Results of the DIA reduction. Left: A ($250^{\prime\prime} \times 500^{\prime\prime}$) 
section of a reference image. Right: a difference image of the same 
section. A number of variables are clearly visible. Plain grey regions
in the difference frame correspond to areas masked during reduction
because of saturation and bad columns in the test image. \label{typi}}
\end{figure}

\newpage
\begin{figure}
\epsscale{1.0}
\plotone{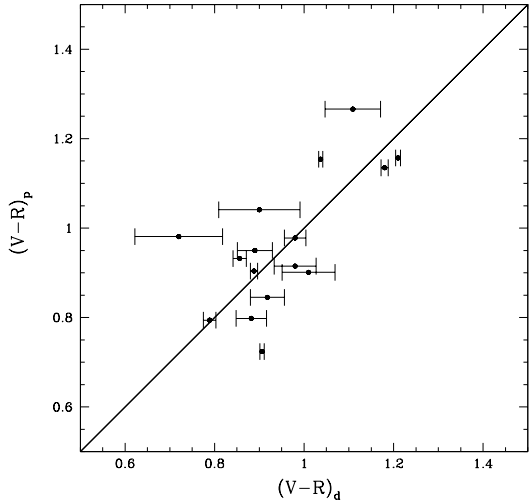}
%\figurenum{11}
\caption{The colours of microlensing sources 
determined from PSF photometry, compared to the colours of the 
same sources determined from difference images. The offsets from the
diagonal line come from blending of the PSF colours.
Errors in the PSF photometry baselines are negligible and so are not shown,
see table \ref{pars}.\label{Macho}}
\end{figure}

\newpage
\begin{figure}
\epsscale{0.8}
\plotone{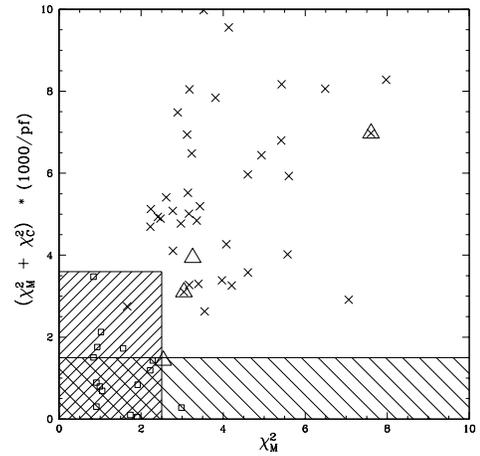}
%\figurenum{12}
\caption{Two sets of cuts performed on microlensing candidates 
  which passed the variability cuts.  Microlensing events with flux at maximum
  amplification is greater than 25000 ADU must lie in the `$\backslash$'
  hatched region.  Events with flux at maximum amplification is less than
  25000 ADU must lie in the `/' hatched region. For specifics see the text.
  Here the $\times$ symbols represent results which fail the constant
  baseline cut, triangles those which fail the V-R colour cut and squares
  the events which pass all cuts. One microlensing event which passes these
  cuts lies outside the region displayed in this figure.\label{cut}}
\end{figure}

\newpage
\begin{figure}
\epsscale{0.85}
\plotone{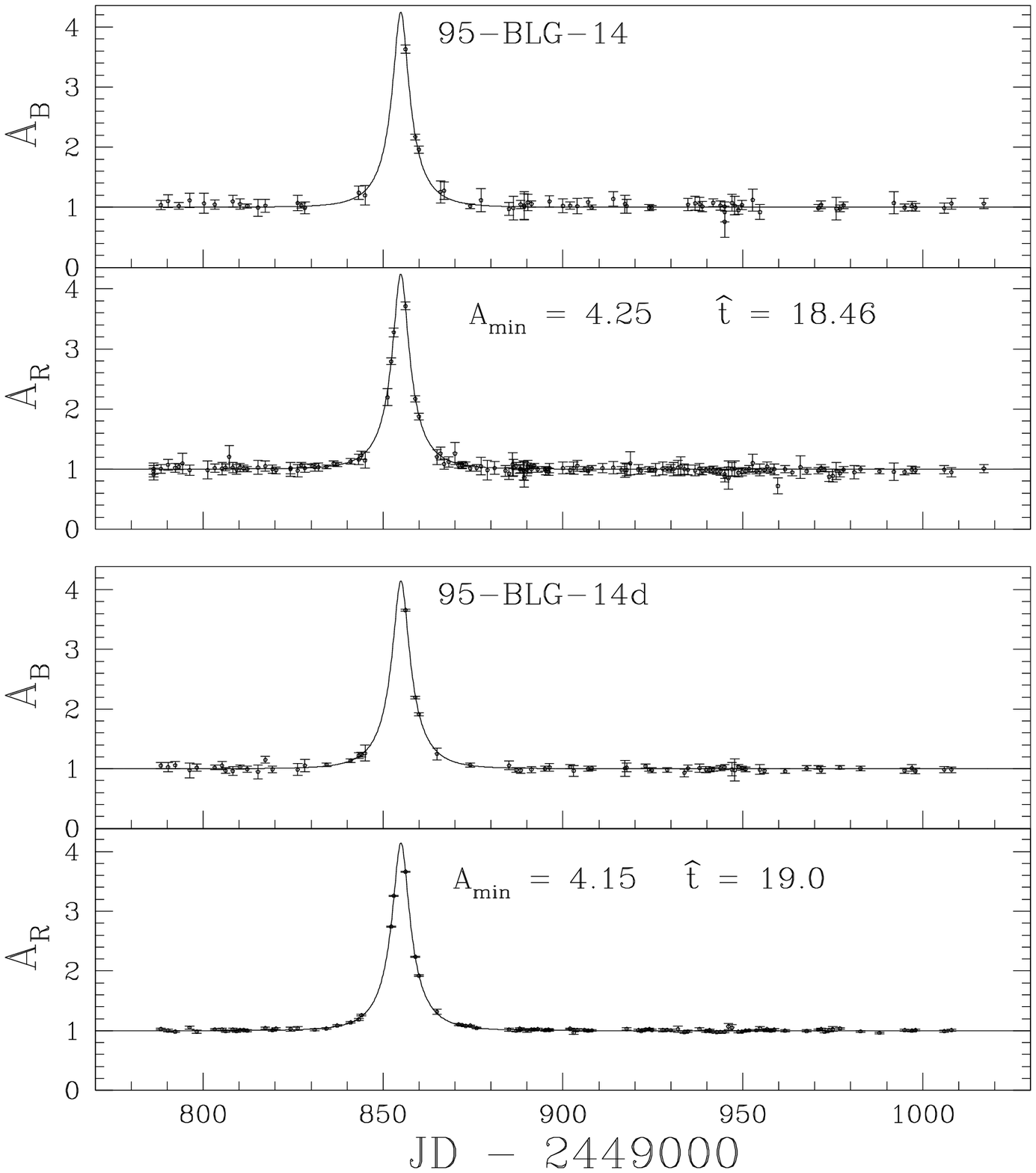}\\
\plotone{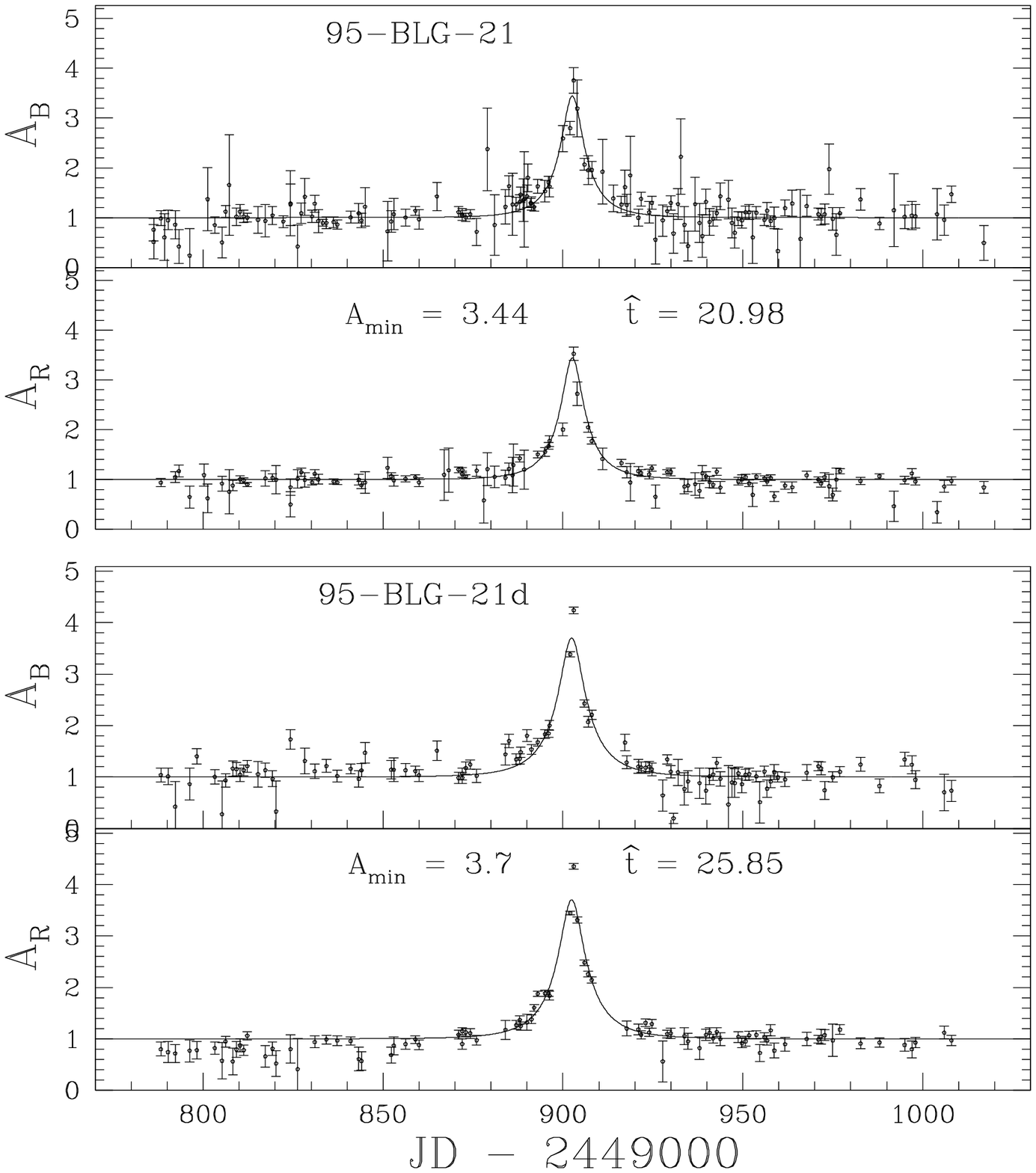}\\
\plotone{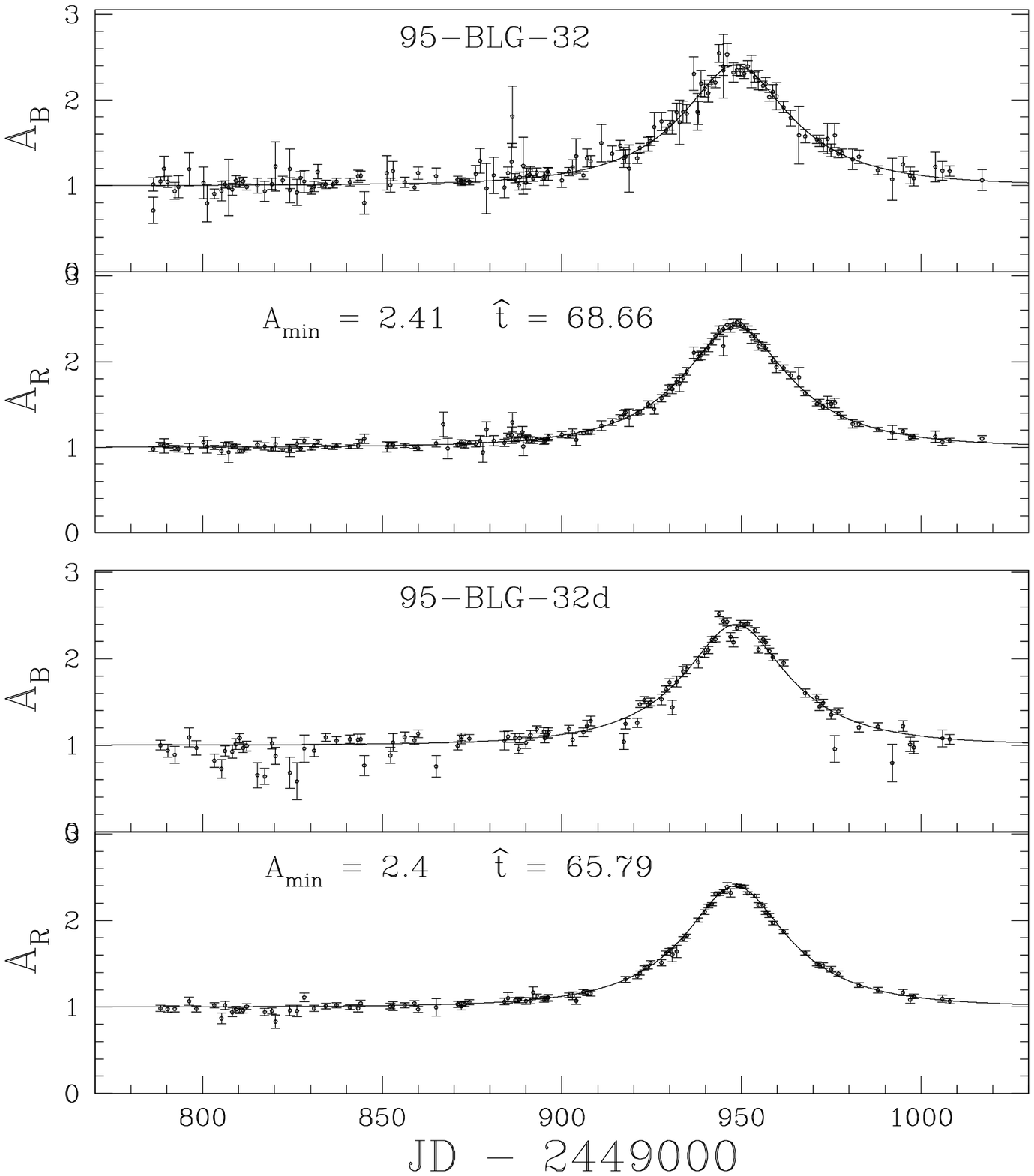}\\
\end{figure}

\newpage
\begin{figure}
\epsscale{0.85}
\plotone{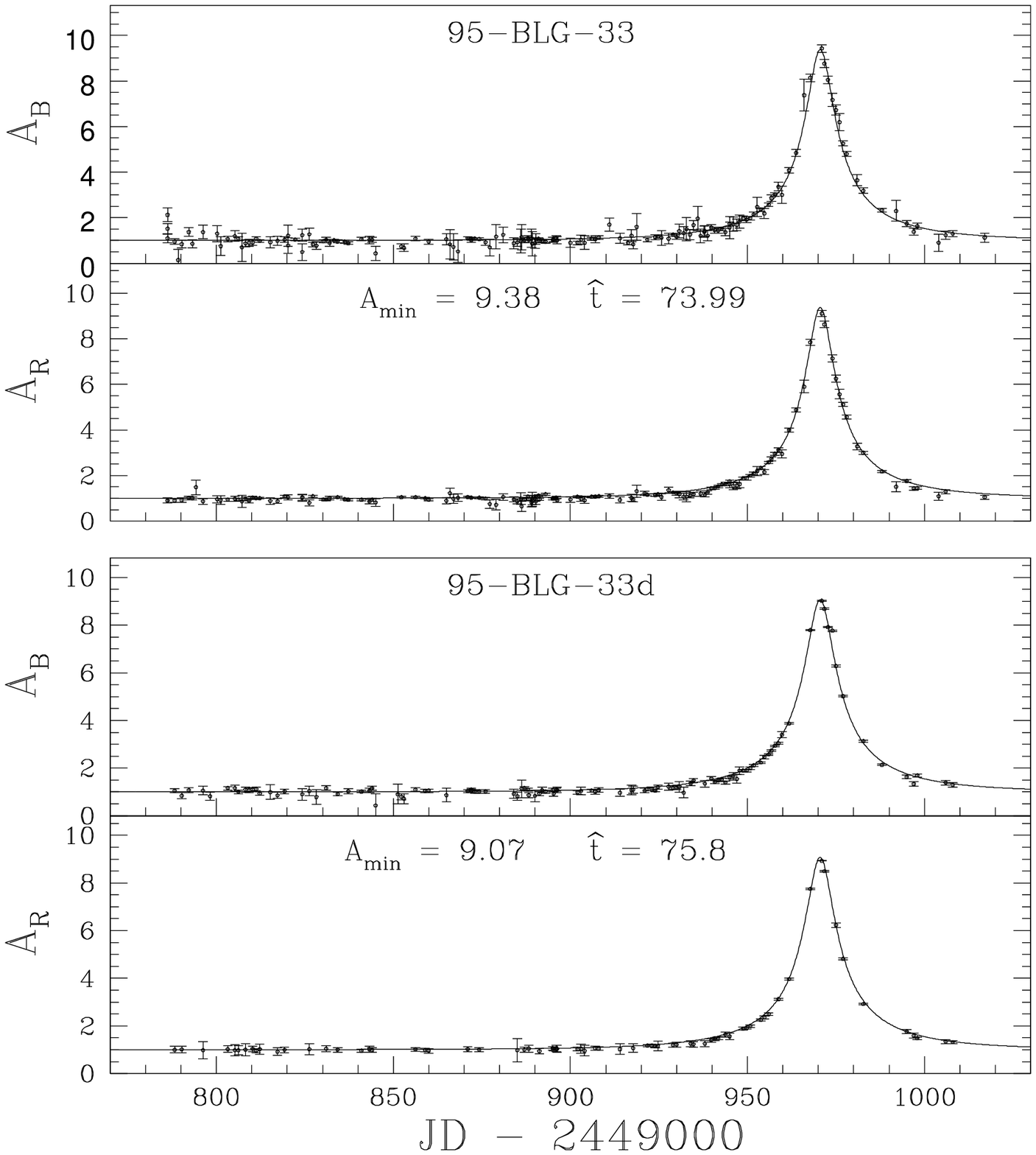}\\
\figurenum{4}
\caption{ Comparisons between the PSF and DIA photometry light curves for 
  candidate classical microlensing events discovered with the Macho Alert
  system and the DIA system.  Data points in the upper panels for each light
  curve come from macho red ($R_{m}$) and blue ($B_{m}$) PSF photometry. 
  Lower panels from red and blue DIA photometry on the same images.\label{Class}}
\end{figure}

\newpage
\begin{figure}
\plotone{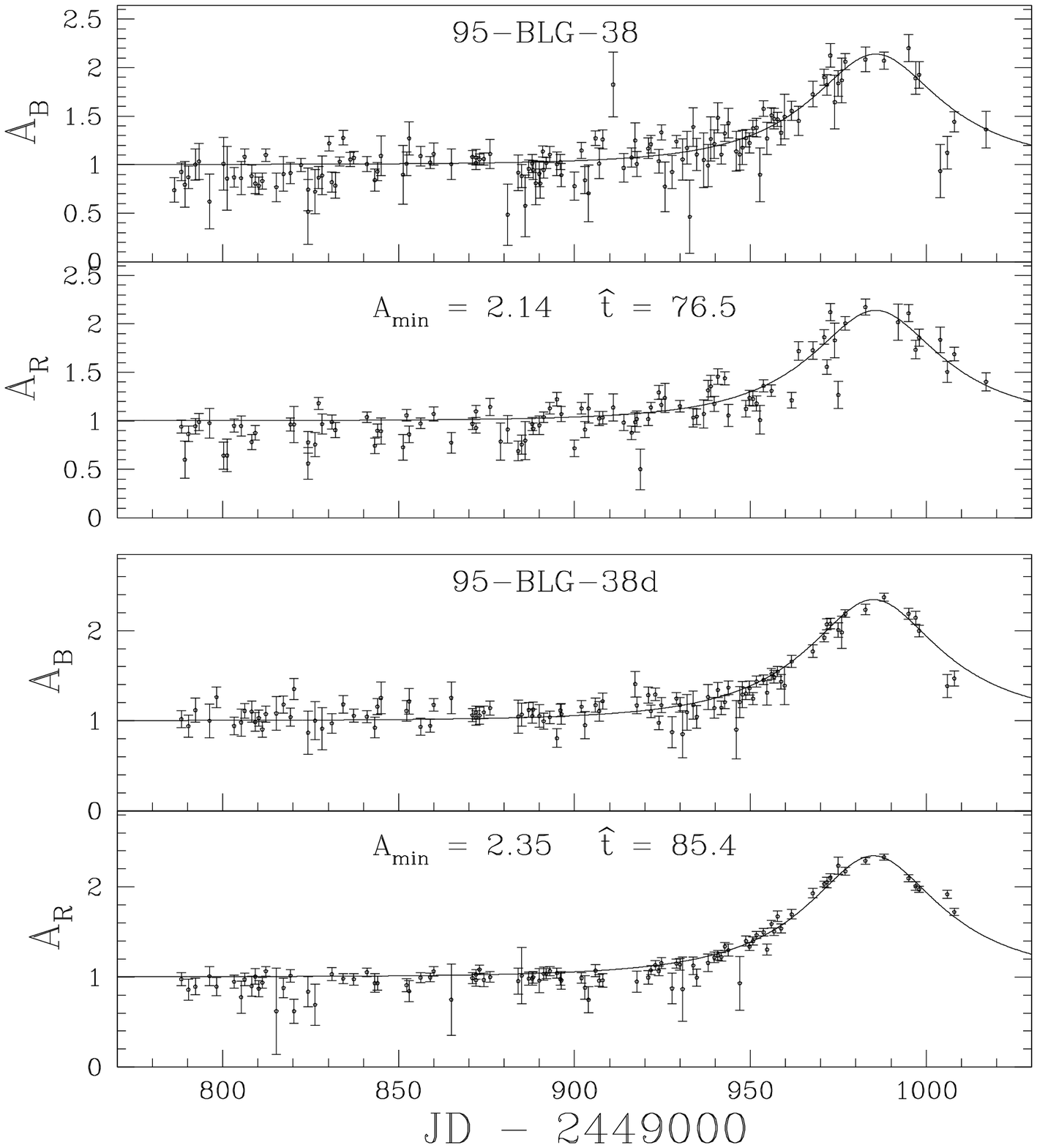}\\
\plotone{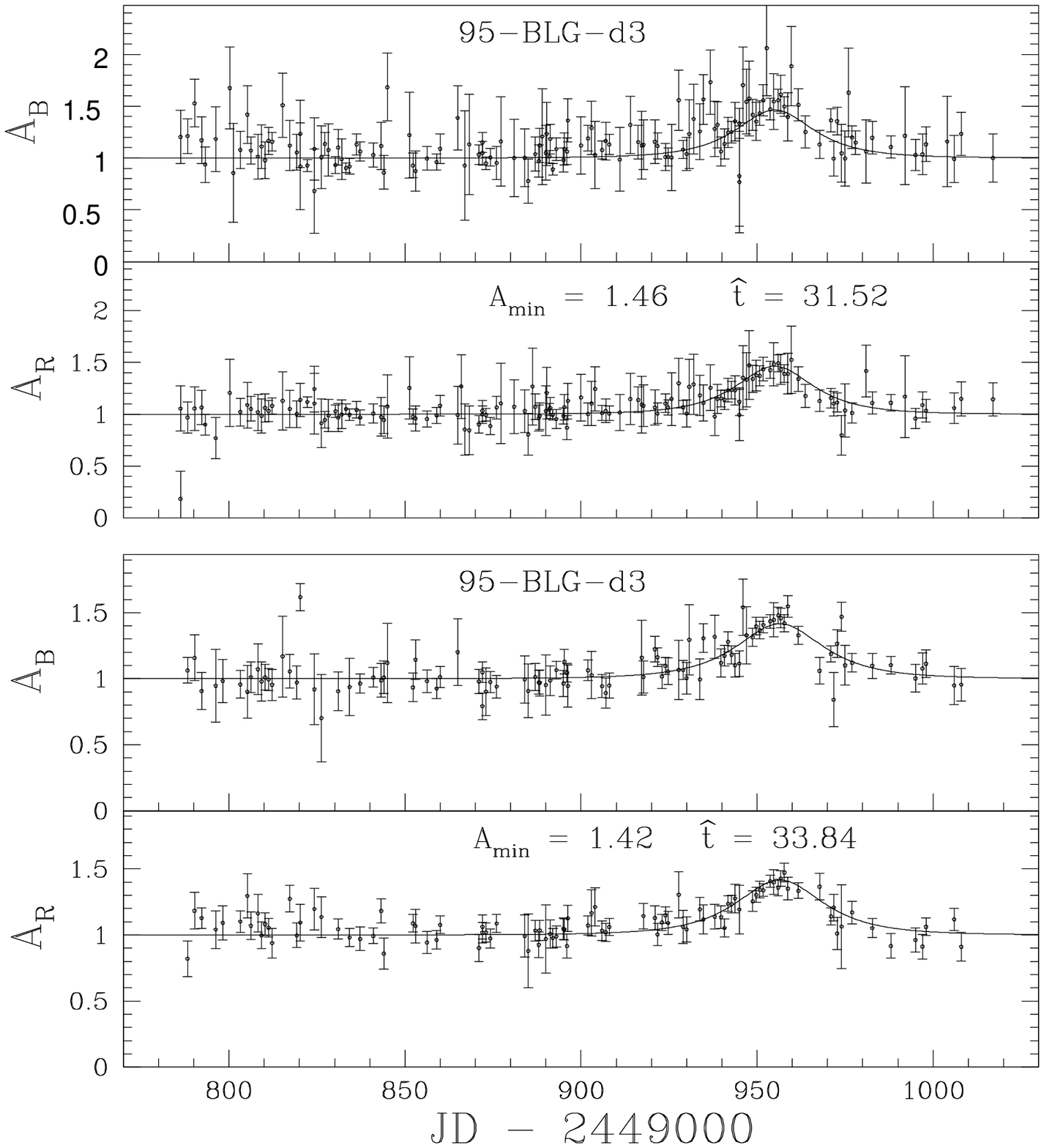}\\
\plotone{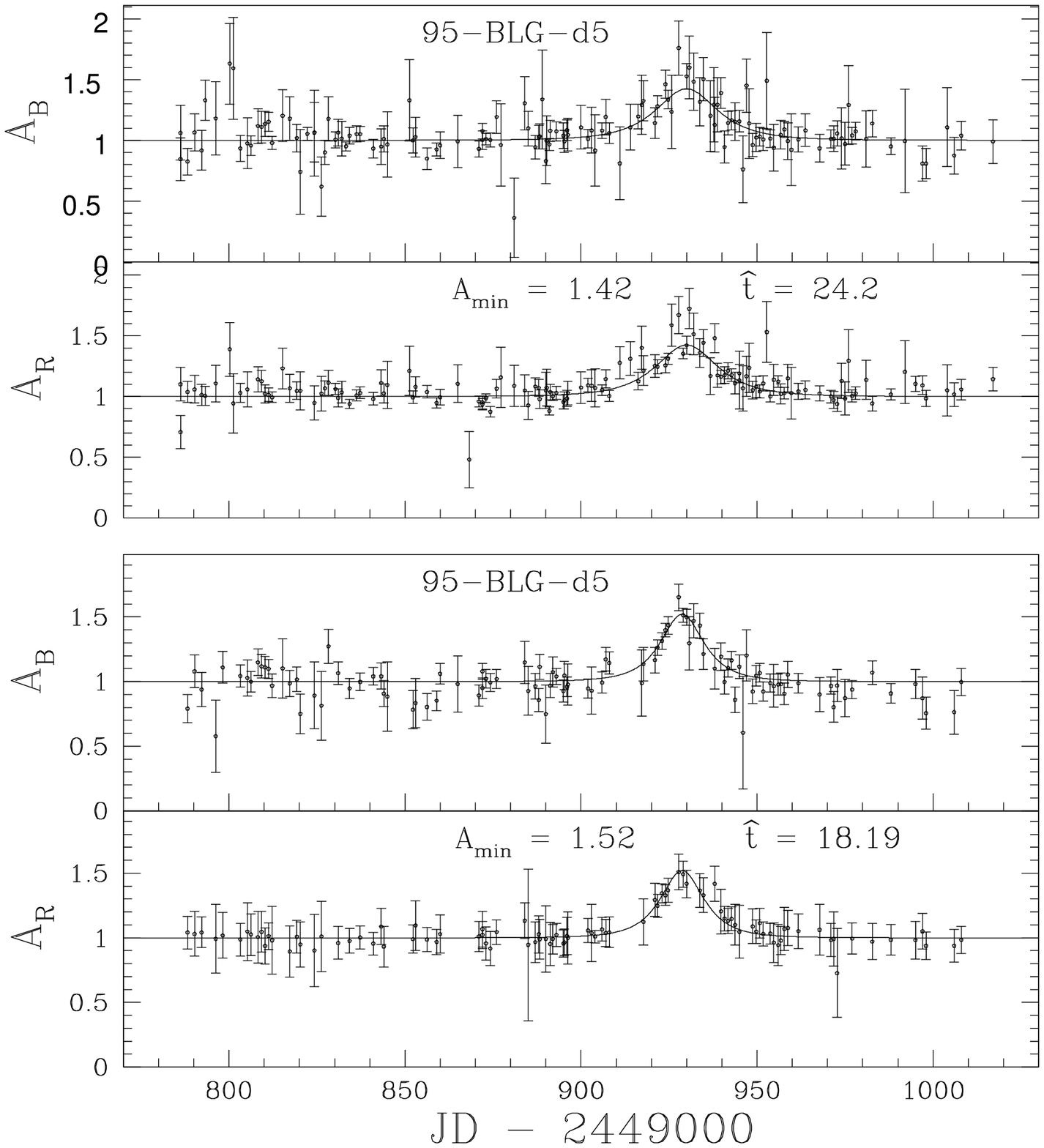}\\
\end{figure}

\newpage
\begin{figure}
\plotone{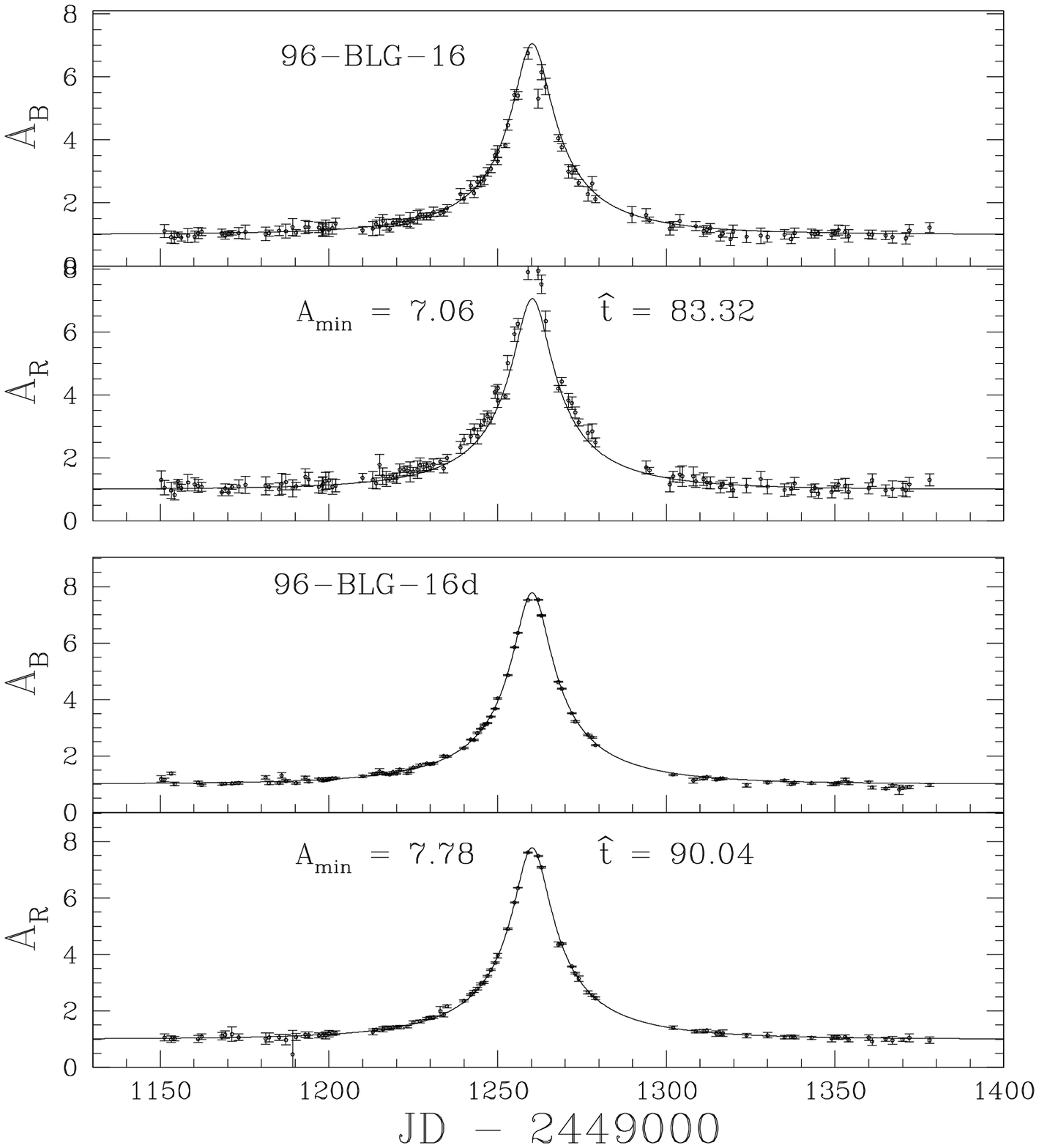}\\
\caption{ {\it continued} \label{Class2}}
\end{figure}

\begin{figure}
\epsscale{0.75}
\plotone{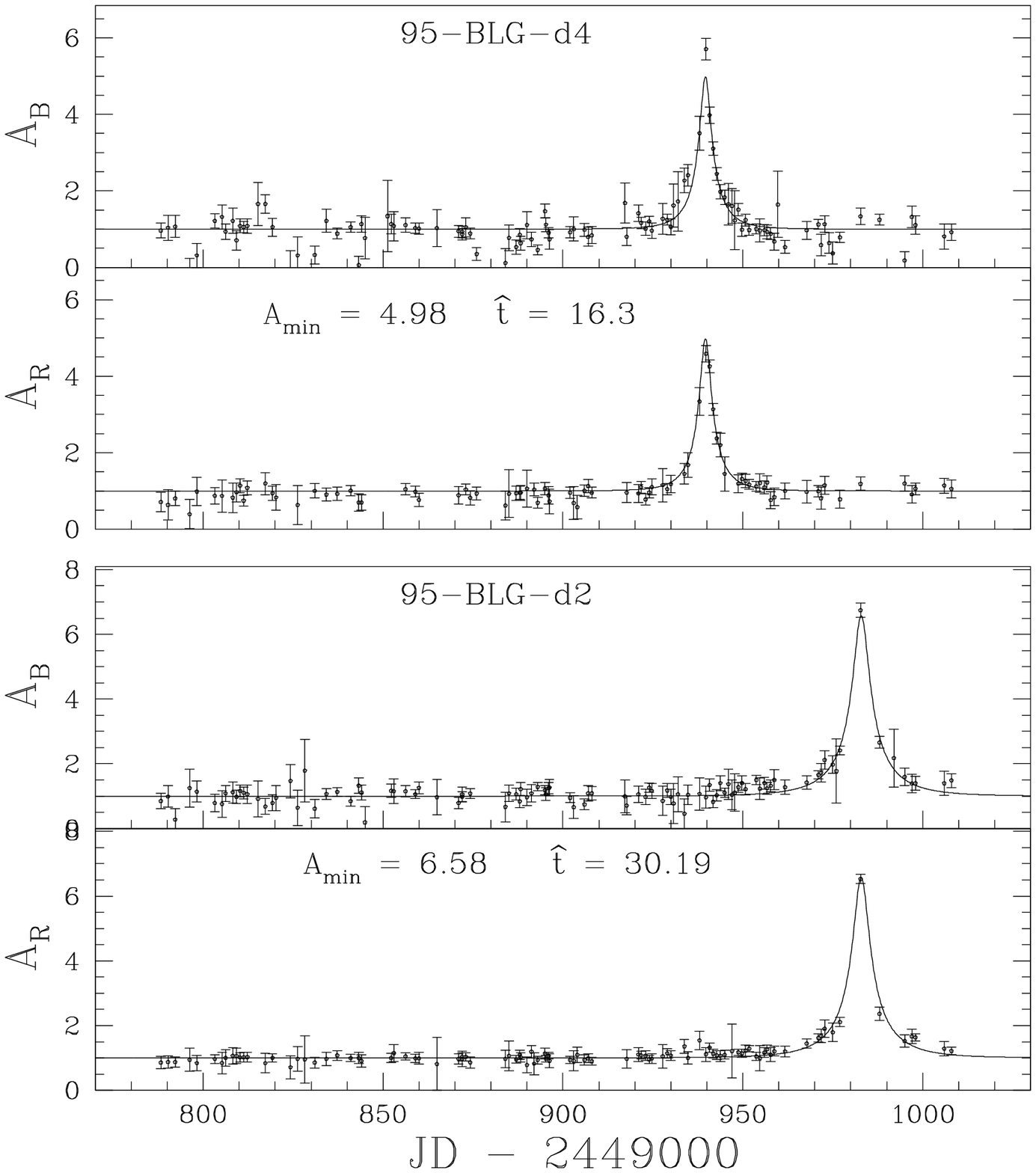}\\
\plotone{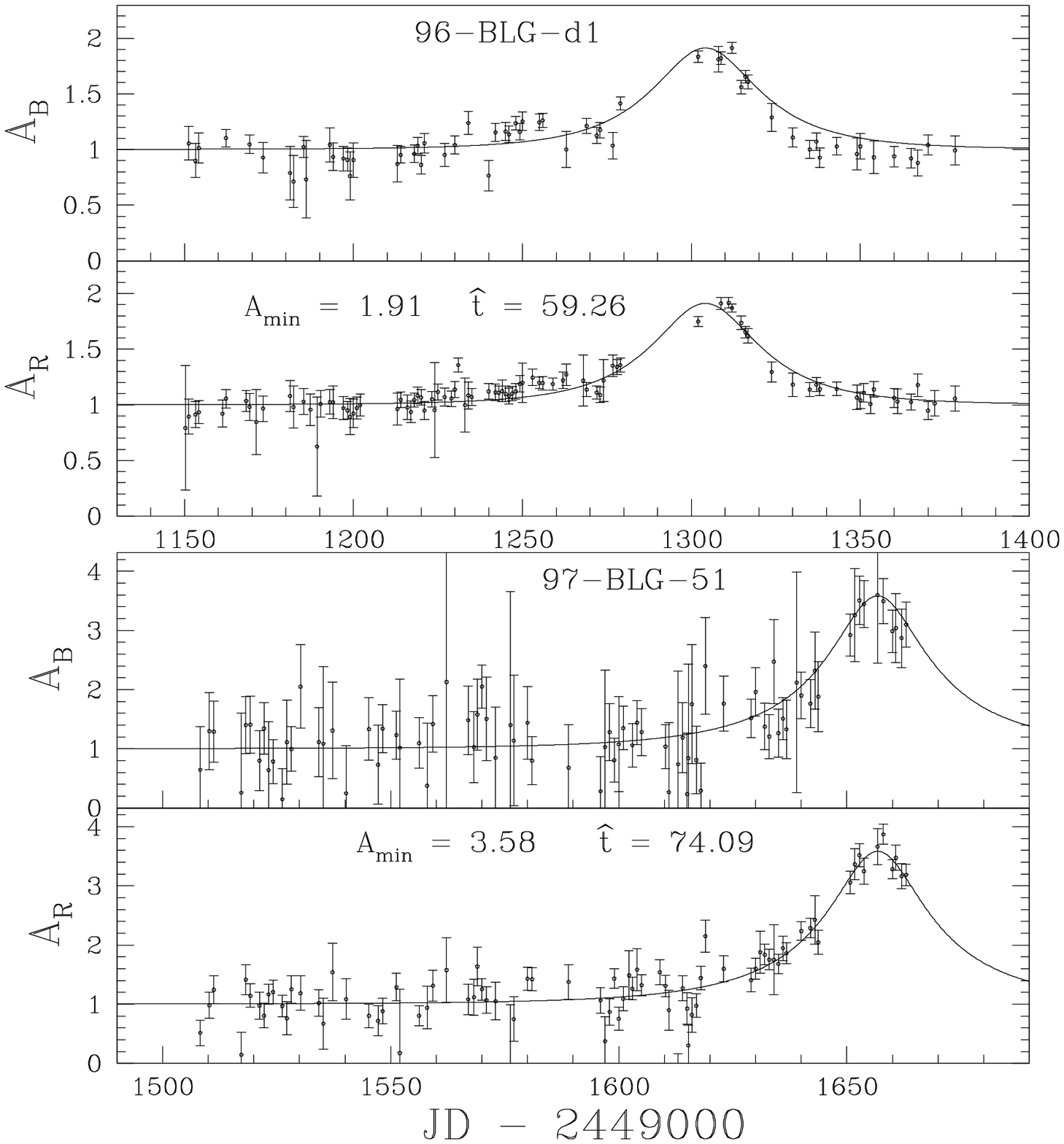}\\
\plotone{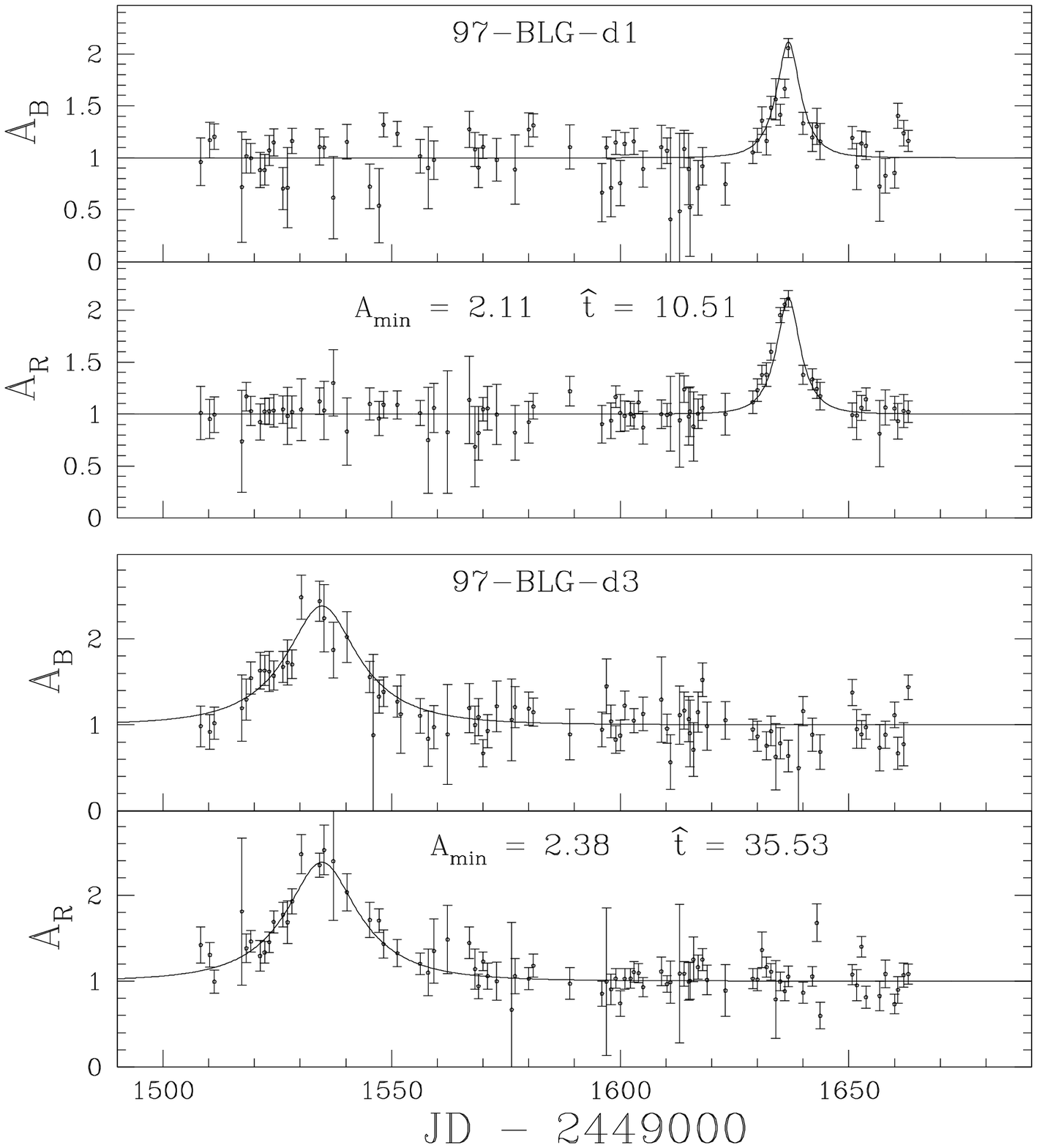}\\
\caption{ This figure shows light curves of candidate
pixel microlensing events discovered with the DIA analysis.
Upper panels are from the blue passband, lower panels are from red
band photometry.\label{Diff}}
\end{figure}

\newpage
\begin{figure}
\epsscale{0.9}
\plotone{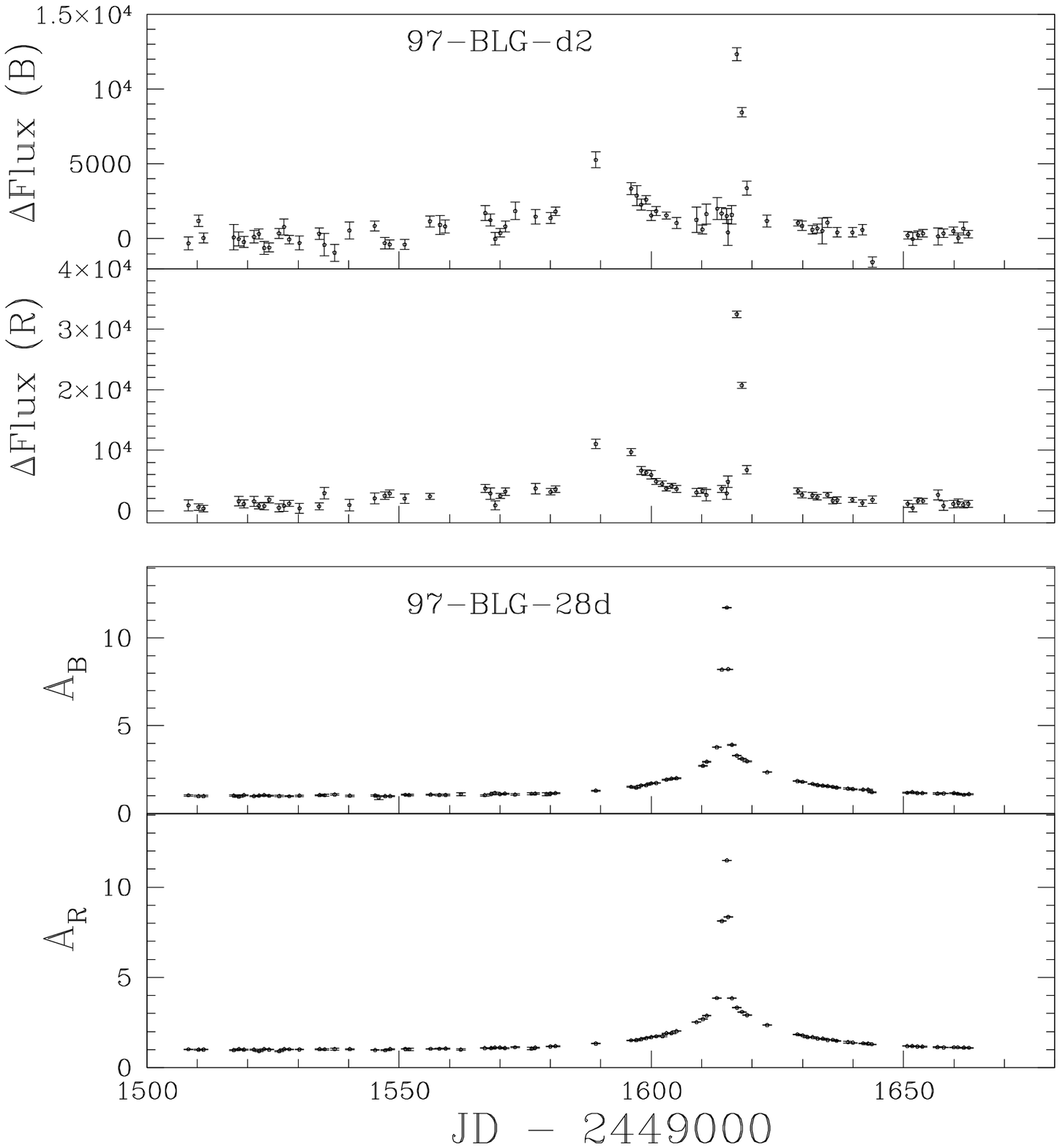}
\caption{ The two binary microlensing events discovered in this analysis.
The top two panels contain a candidate binary pixel lensing event. 
Here the baseline flux is unknown so difference flux is plotted instead of A.
The lower panels contain an event with a cusp crossing discovered by the 
Macho alert system. The fits for these events will be presented in later 
work (\cite{alc99b}).\label{Unfit}}
\end{figure}

% TABLE1.TEX -- Stuff.

%\begin{landscape}
\begin{deluxetable}{lrrrrrrrrr}
%\notetoeditor{Table \ref{pars} should be rotated and occupy an entire page
%if too big.}
\tablewidth{0pt}
\footnotesize
\tablecaption{Parameters of Events.\label{pars}}

\tablehead{
\colhead{Event} & \colhead{R.A. $(2000)$}  & \colhead{Decl. $(2000)$}  &
\colhead{V} & \colhead{(V-R)$_{p}$} & \colhead{(V-R)$_{d}$} &  \colhead{(V-R)$_{rr}$} &
\colhead{(V-R)$_{c0}$} & \colhead{S ($^{\prime\prime}$)} & \colhead{Alert}\\
\colhead{(1)} & \colhead{(2)} & \colhead{(3)} & 
\colhead{(4)} & \colhead{(5)} & \colhead{(6)} &
\colhead{(7)} & \colhead{(8)} & \colhead{(9)} &
\colhead{(10)}}

\startdata
95 BLG 33 & 18 01 57.1  & $-28$ 08 06 & 19.14 & 1.152(3) & 1.037(5)  & 0.89(5) & 0.397(50) & 0.16(0.2) & y\nl 
95 BLG d2 & 18 01 13.8  & $-28$ 01 25 & 20.13 & 1.041(6) & 0.900(91) & 0.80(5) & 0.350(60) & 0.53(0.3) & n\nl
97 BLG 28 & 18 00 33.8  & $-28$ 01 10 & 18.10 & 1.155(1) & 1.210(5)  & 0.83(5) & 0.640(50) & 0.12(0.2) & y\nl
97 BLG 51 & 18 00 39.3  & $-27$ 58 35 & 20.56 & 0.915(7) & 0.980(47) & 0.74(6) & 0.49(10)  & 0.79(0.5) & y\nl
96 BLG d1 & 18 00 36.1  & $-27$ 58 30 & 19.41 & 0.901(4) & 1.010(59) & 0.83(5) & 0.430(50) & 0.62(0.3) & n\nl
95 BLG 32 & 18 00 49.9  & $-27$ 58 44 & 18.94 & 1.135(3) & 1.180(8)  & 0.83(5) & 0.600(50) & 0.14(0.2) & y\nl
96 BLG 16 & 18 01 09.0  & $-27$ 57 59 & 18.34 & 0.724(4) & 0.906(5)  & 0.74(10) & 0.44(10) & 0.18(0.2) & y\nl 
95 BLG 38 & 17 59 41.9  & $-28$ 12 10 & 19.48 & 0.978(4) & 0.980(24) & 0.73(10) & 0.50(10) & 0.12(0.2) & y\nl
95 BLG 21 & 17 59 42.1  & $-28$ 08 41 & 20.13 & 0.950(7) & 0.890(39) & 0.73(10) & 0.41(10) & 0.87(0.4) & y\nl
95 BLG d3 & 18 00 35.6  & $-28$ 38 14 & 19.13 & 0.794(4) & 0.789(14) & 0.70(3) & 0.339(40) & 0.16(0.2) & n\nl
97 BLG d2 & 18 00 39.6  & $-28$ 34 43 & \nodata\tablenotemark{a} & \nodata\tablenotemark{a}   & 0.876(33) & 0.70(3) & 0.426(50) & 0.95(0.2) & n\nl
97 BLG d3 & 18 00 36.1  & $-28$ 32 34 & 18.38 & 0.932(5) & 0.860(15) & 0.67(4) & 0.440(40) & 0.78(0.2) & n\nl
95 BLG d4 & 18 00 34.4  & $-28$ 33 30 & 19.92 & 0.798(7) & 0.882(34) & 0.67(4) & 0.462(60) & 0.90(0.2) & n\nl
95 BLG d5 & 18 01 47.1  & $-28$ 21 26 & 19.28 & 0.845(5) & 0.918(38) & 0.71(3) & 0.458(60) & 0.11(0.3) & n\nl
97 BLG d1 & 18 01 30.9  & $-28$ 32 16 & 19.62 & 0.981(5) & 0.720(98) & 0.75(8) & 0.22(13) & 0.89(0.4) & n\nl
95 BLG 14 & 18 01 26.3  & $-28$ 31 14 & 17.67 & 0.904(2) & 0.888(8)  & 0.75(8) & 0.388(80) & 0.06(0.2) & y\nl
\tablenotetext{a}{ No blue photometry available for this object}
\tablecomments{Col. (4) shows the V magnitude of the nearest Macho object. Cols. (5) and
(6) show the colour from the PSF photometry and difference images respectively.
Col. (7) gives the colour of the nearest RR Lyrae stars used for differential reddening correction.
Col. (8) is the reddening corrected cousins colour. Col. (9) is the separation of the difference 
image centroid and the nearest photometered object. Col. (10) denotes whether the event was alerted
on from the photometry. Cols. (4)$-$(8) include the 
$1\sigma$ associated errors in the last one or two significant digits 
shown in parentheses.}
\enddata
\end{deluxetable}
%\end{landscape}

% TABLE2.TEX -- table 2.

%\clearpage
%\begin{landscape}
\begin{deluxetable}{lrrrrrrrrrr}
\notetoeditor{Table \ref{fit} should be rotated and occupy an entire page
if too big.}
\scriptsize
\tablecaption{Parameters of Classical Event Fits.\label{fit}}
\tablewidth{0pt}
\tablehead{
\colhead{Event} & \colhead{t$_{max}$} & \colhead{A$_{min}$} & \colhead{$\hat t_{min,d}$} &
\colhead{$\hat t_{min,p}$} & \colhead{$\chi^{2}_{d}$/$N_{deg}$} & \colhead{$\chi^{2}_{p}$/$N_{deg}$} &
\colhead{ErrR$_{d}$\%} & \colhead{ErrR$_{p}$\%} & \colhead{ErrB$_{d}$\%} & 
\colhead{ErrB$_{p}$\%} \\
\colhead{(1)} & \colhead{(2)} & \colhead{(3)} & 
\colhead{(4)} & \colhead{(5)} & \colhead{(6)} &
\colhead{(7)} & \colhead{(8)} & \colhead{(9)} &
\colhead{(10)} & \colhead{(11)}}
\startdata
95 BLG 14 & 854.98(1) & 4.23 & 19.3 & 18.5 & 0.91 &  0.46 & 1.12 & 2.18 & 1.97 & 2.97\nl
95 BLG 21 & 902.47(7) & 3.70 & 25.9 &  21.0 & 1.30  & 1.54 & 8.58 & 7.29 & 11.3 & 11.5\nl
95 BLG 32 & 948.4(1) & 2.40 & 65.8  & 68.7 & 2.22  & 1.03 & 1.51 & 1.78 & 4.73 & 3.98\nl
95 BLG 33 & 970.48(19) & 9.07 & 75.8 & 74.0 & 2.48 & 1.22 & 2.46 & 5.69 & 5.60 & 6.05 \nl
95 BLG 38 & 985.00(31) & 2.35 & 85.4 & 76.5  & 1.25  & 2.15 & 4.27 & 8.19 & 7.04 & 8.05 \nl
96 BLG 16 & 1260.23(1) & 7.78 & 90.0 & 83.3  & 2.96  & 0.93 & 2.69 & 7.76 & 4.43 & 6.85\nl
95 BLG d3 & 956.4(8) & 1.42 & 33.8 & 31.5  & 0.99 & 0.56 & 4.96 & 4.36 & 5.36 & 8.13\nl
95 BLG d5 & 928.70(47) & 1.52  & 18.2 & 24.2 & 0.48 & 0.87 & 3.83 & 4.56 & 6.07 & 6.21\nl
\tablecomments{Cols. (2) \& (3) give difference image photometry microlensing
fit values for amplification and $t_{0}$ (microlensing time ($JD-2449000$) of closest approach), 
with the formal $1\sigma$ fit error in the last one or two significant digits shown in
parentheses. Cols. (4) to (7) give the minimum event time scale $\hat t$ and $\chi^{2}$ values 
obtained from fits to difference image photometry and PSF photometry.
Cols. (8) to (11) give the median residual scatter of the photometry points about the fits
for Blue and Red difference images ($B_{d}$, $R_{d}$) and PSF photometry ($B_{p}$, $R_{p}$). 
Fits assume PSF photometry baselines.}
\enddata
\end{deluxetable}
%\end{landscape}

% TABLE2.TEX -- table 3.

\clearpage
\begin{deluxetable}{crrrrrr}
\footnotesize
\tablecaption{Parameters of Pixel Lensing Event Fits.\label{fit2}}
\tablewidth{0pt}
\tablehead{
\colhead{Event} & \colhead{t$_{max}$} & \colhead{A$_{min}$} & 
\colhead{$\hat t_{min}$} & \colhead{$\chi^{2}/N_{deg}$} & 
\colhead{ErrR$_{d}$\%} & \colhead{ErrB$_{d}$\%}\\
\colhead{(1)} & \colhead{(2)} & \colhead{(3)} & 
\colhead{(4)} & \colhead{(5)} & \colhead{(6)} &
\colhead{(7)}
}
\startdata
97 BLG 51 & 1656.77$(19)$ & 3.58 & 74.1 & 1.41 & 22.9 & 31.3 \nl
95 BLG d2 & 982.9$(1)$  & 6.58 & 30.19 &  0.66 & 7.29 & 11.7\nl
95 BLG d4 & 939.65$(6)$ & 3.14 & 10.34 & 1.04  & 8.61 & 18.6 \nl
96 BLG d1 & 1304.1$(6)$ & 1.91 & 59.3 & 0.93 & 4.22 & 8.74\nl
97 BLG d1 & 1636.79$(16)$ & 2.11 & 10.5  & 0.84 & 7.65 & 13.5\nl
97 BLG d3 & 1534.8$(6)$ & 2.38 & 35.53 & 1.05 & 9.59 & 14.5\nl
\tablecomments{ Cols. (2) to (5) are difference image photometry
microlensing fit values. Col. (2) gives the microlensing times (JD$-2449000$)
of closest approach, with the formal $1\sigma$ fit error in the last one or
two significant digits shown in parentheses. Col. (3) presents the minimum
amplification of the sources. Col. (4) gives the minimum event time
scales. Col. (5) gives the reduced $\chi^{2}$ values of the
microlensing fits. Cols. (6) and (7) give the median residual scatter of
the photometry points about the microlensing fits.}
\enddata
\end{deluxetable}

\end{document}